# Physics-Guided Sequence Modeling for Fast Simulation and Design Exploration of 2D Memristive Devices

*Benjamin Spetzler[1,*], Elizaveta Spetzler[2], Saba Zamankhani[3], Dilara Abdel[4], Patricio Farrell[4], Kai-Uwe Sattler[3], Martin Ziegler[1]*

[1] B. Spetzler, M. Ziegler
Energy Materials and Devices, Department of Materials Science, Faculty of Engineering, Kiel University, 24143 Kiel, Germany, E-mail: besp@tf.uni-kiel.de

[2] E. Spetzler
Nanoscale Magnetic Materials – Magnetic Domains, Department of Materials Science, Faculty of Engineering, Kiel University, 24143 Kiel, Germany

[3] S. Zamankhani, K.-U. Sattler
Databases and Information Systems, Department of Computer Science and Automation, TU Ilmenau, Ilmenau 98693, Germany

[4] D. Abdel, P. Farrell
Numerical Methods for Innovative Semiconductor Devices, Weierstrass Institute for Applied Analysis and Stochastics (WIAS), Mohrenstr. 39, 10117 Berlin, Germany

**Abstract**

Modeling hysteretic switching dynamics in memristive devices is computationally demanding due to coupled ionic and electronic transport processes. This challenge is particularly relevant for emerging two-dimensional (2D) devices, which feature high-dimensional design spaces that remain largely unexplored. We introduce a physics-guided modeling framework that integrates high-fidelity finite-volume (FV) charge transport simulations with a long short-term memory (LSTM) artificial neural network (ANN) to predict dynamic current-voltage behavior. Trained on physically grounded simulation data, the ANN surrogate achieves more than four orders of magnitude speedup compared to the FV model, while maintaining direct access to physically meaningful input parameters and high accuracy with typical normalized errors <1%. This enables iterative tasks that were previously computationally prohibitive, including inverse modeling from experimental data, design space exploration via metric mapping and sensitivity analysis, as well as constrained multi-objective design optimization. Importantly, the framework preserves physical interpretability via access to detailed spatial dynamics, including carrier densities, vacancy distributions, and electrostatic potentials, through a direct link to the underlying FV model. Our approach establishes a scalable framework for efficient exploration, interpretation, and model-driven design of emerging 2D memristive and neuromorphic devices.

## 1. INTRODUCTION

Emerging electronic devices increasingly rely on dynamic, nonlinear, and history-dependent transport processes that are challenging to model and optimize. Systems such as electrolyte-gated transistors [1,2], ion-gel-based transistors [2–4] , and vacancy-mediated memristive devices [5,6] exhibit complex hysteresis phenomena arising from coupled electronic and ionic dynamics. Accurately capturing these behaviors is essential for advancing neuromorphic computing, memory technologies, and adaptive electronics. However, conventional numerical models often remain computationally prohibitive for large-scale parameter exploration, inverse modeling, or device optimization [7–9]. This motivates the development of modeling strategies that combine physical knowledge with efficient predictive modeling.

Vacancy-mediated memristive devices based on two-dimensional materials provide a representative example of such systems [5,6]. Their atomic-scale thickness, surface sensitivity, tunable electronic properties, and the emergence of moiré heterostructures offer a largely unexplored design space for tuning memristive charge transport phenomena [10–15]. Particularly, lateral device architectures integrating additional gate electrodes have led to the development of memtransistors, multi-terminal devices that combine nonvolatile resistive switching with the electric field tunability of transistors [16–25]. These devices permit electrostatic modulation of memristive hysteresis, supporting high linearity and symmetry in synaptic responses, tunable learning rates, heterosynaptic plasticity, and many programmable states [16,17,19,26–28]. Recent studies on memtransistors based on transition metal dichalcogenides (TMDCs) such as $MoS_2$ have demonstrated sub-micron channel lengths, sub-1 V operation, and low switching energies, illustrating their promise for scalable neuromorphic systems [17,26,29,30].

Advancing the performance and integration of these devices requires a deeper understanding of the physical origins of memristive behavior, particularly the influence of material properties, device structure, and operating conditions on dynamic switching behavior and hysteresis. Although experiments have revealed promising device characteristics [30–34], this vast design space remains challenging to explore empirically [8,20]. Moreover, memristive switching in 2D materials can involve a complex interplay of physical mechanisms, including ionic motion, interface and grain boundary effects, trap state dynamics, and gate-induced modulation [5,31]. Even when focusing on a single material such as $MoS_2$, the most widely studied TMDC, the relevant material properties can vary over orders of magnitude due to variations in microstructure, defect concentration, strain, interface quality, and processing conditions [32–37]. Existing computational approaches struggle to capture this full complexity [8].

A first step toward addressing this challenge is to develop tractable, physically grounded models that can reproduce key features of device behavior to guide design and optimization.

To date, relatively few computational models have been presented to study switching in 2D memristive devices and memtransistors [25,38–41]. Compact models use simplified physics-inspired descriptions to emulate I-V characteristics for electrical equivalent circuit simulations [38]. They are suitable for system level modeling but do not directly reflect microscopic transport phenomena. Kinetic Monte Carlo models [39,40] incorporate stochastic defect migration limited to nanoscale volumes, often omitting semiconductor-metal interfaces and detailed semiconductor transport. More recently, quasi-static [25] and fully transient [41,42] charge transport models have been proposed that solve coupled partial differential equations (PDEs) describing charge and mass transport in lateral geometries with $MoS_2$ channels. This approach captures the spatial and temporal dynamics of vacancy dominated switching, providing some direct links between the design parameters, physical mechanisms, and device performance [8,43]. However, the complexity of the charge transport mechanisms and the high-dimensionality of the parameter space lead to prohibitive computational costs for highly iterative tasks such as parameter fitting, design-space exploration, and optimization [8].

To approach these challenges, we introduce a physics-guided model that combines high-throughput charge transport simulations with artificial neural networks (ANNs) for fast, interpretable prediction of memristive behavior. The approach is particularly suited to emerging device platforms, such as 2D-material-based memristive devices, where the underlying physics is complex, the design space is high-dimensional, and experimental data may be limited or costly to obtain. By using simulation data as a high-quality training set, the model inherits physical realism and interpretability from the simulation while gaining orders-of-magnitude speedup through a trained neural network surrogate. While physics-informed and machine learning accelerated modeling has gained traction in computational materials science for materials and physics discovery [44–46], including PDE systems [47–50], applications to memristive devices remain rare. Existing physics-informed and data-driven models are primarily limited to compact models based on equivalent circuits for metal-oxide systems [51–55].

In this work, we train long short-term memory (LSTM) networks to predict dynamic I-V curves from physically interpretable input features, derived directly from high-throughput, finite volume charge transport simulations, enabling rapid surrogate modeling across an extended range of material and device parameters. We validate the model against experimental data, quantify its accuracy, and apply it to three general use cases enabled by the ANN model: (i) inverse modeling for experimental parameter extraction, (ii) fast design space exploration via metric mapping and global sensitivity analysis, and (iii) constrained performance optimization (Figure 1). Together, these capabilities establish a scalable modeling framework that complements high-fidelity simulations and supports data-driven analysis and design of emerging memristive devices.

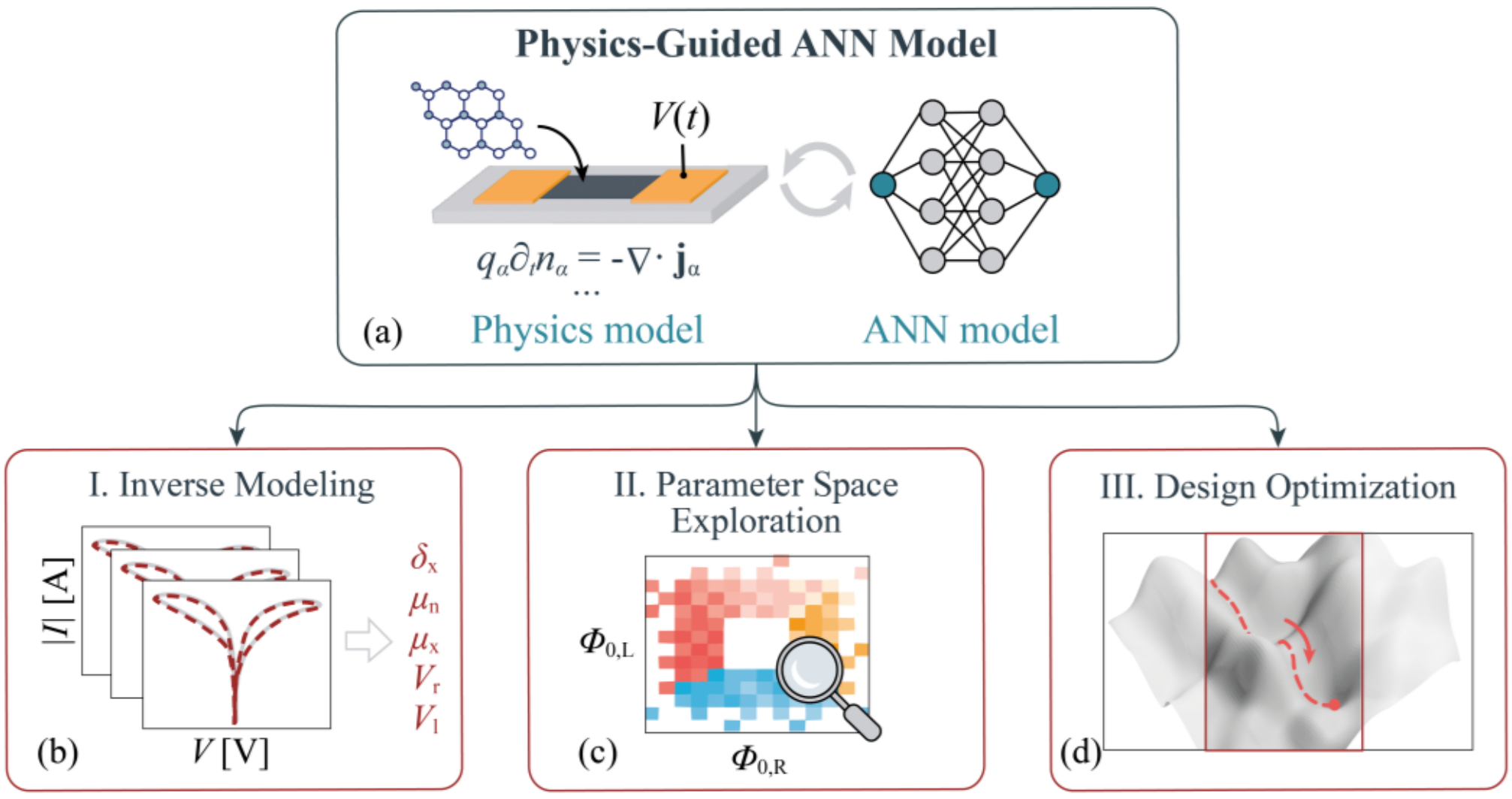

**Figure 1. Illustration of the proposed physics-guided ANN for model exploration and device design. a** The proposed framework combines high-fidelity charge transport simulations based on a finite-volume discretization of a drift-diffusion system of coupled PDEs with a long short-term memory (LSTM) artificial neural network (ANN) to model time-dependent current-voltage (I-V) behavior in 2D memristive devices. **b** The trained surrogate model enables fast inverse modeling by fitting experimental I-V data to extract physically meaningful device parameters. **c** The ANN accelerates exploration of high-dimensional parameter spaces, enabling the mapping and global analysis of device metrics such as hysteresis symmetry and on-off ratio across a range of material and interface parameters. **d** The model also supports multivariate constrained performance optimization by enabling efficient search for parameter combinations that yield targeted device characteristics.

# 3 RESULTS

## 3.1 Physics-Guided ANN Model

To simulate vacancy-dominated hysteretic switching in two-dimensional (2D) memristive devices, exemplified here by lateral $MoS_2$ structures (Figure 2a,b), we developed a modeling framework that combines finite-volume (FV) charge transport simulations with an LSTM-based artificial neural network (ANN) surrogate model.

**Charge transport model.** The charge transport model solves the fully coupled transient drift-diffusion equations for electrons, holes, and positively charged vacancies, together with Poisson's equation, using quasi-Fermi potentials and the electrostatic potential as unknowns. It accounts for the nonlinear dynamics of all carriers and the distinct physical nature of vacancies, including their limited maximum density imposed by the material structure. To capture the highly anisotropic mobilities in layered materials [37,56], migration is constrained to the channel direction. In contrast to most conventional models, we account for the local electric-field dependence of Schottky barriers at metal-semiconductor interfaces by self-consistently solving for an additional auxiliary potential describing image-charge-induced Schottky barrier lowering (SBL). Further, the equations are discretized using the Voronoi finite volume method, ensuring local flux conservation, thermodynamic consistency, and physically meaningful carrier densities [57,58].

Figure 2a,b illustrates the general device geometry with two metal electrodes positioned at either end of the semiconducting channel. For all simulations, a time-dependent triangular voltage sweep $V(t)$ is applied between the electrodes (Figure 2a,c), with the left contact grounded. Due to their significantly lower mobility, sulfur vacancies respond with a delay to the applied electric field. As the voltage is swept, the vacancies gradually accumulate near

one contact and deplete near the other, dynamically modifying the local space charge, altering carrier concentrations, and shifting the Schottky barrier heights (Figure 2d). The delayed redistribution produces nonlinear I-V characteristics that can exhibit pronounced hysteresis depending on the input parameters (Figure 2e). This mechanism is supported by prior experimental and theoretical studies on $MoS_2$ [42,59,60] and reflects the prevailing understanding of memristive switching in this system.

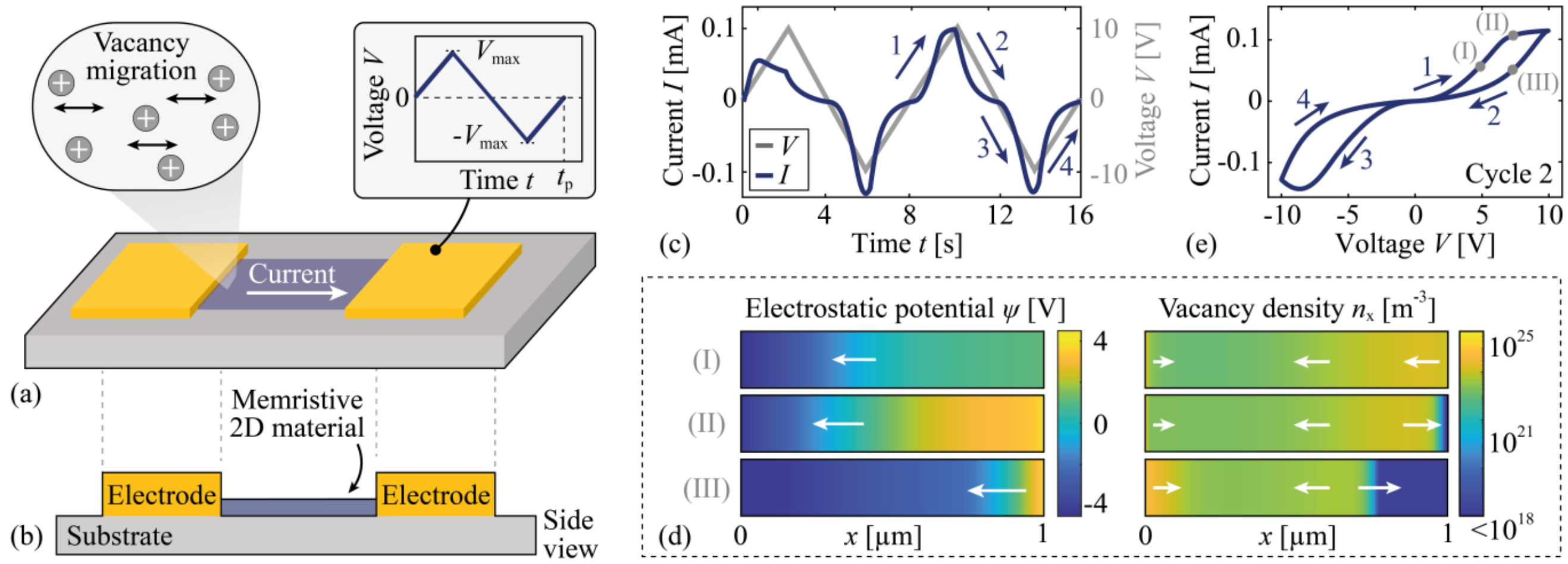

**Figure 2. Device geometry, voltage protocol, and simulated switching behavior.**

**a** Illustration of a lateral memristive device consisting of a two-dimensional memristive material contacted by two electrodes on a substrate. A voltage is applied to the right electrode following a piecewise linear function of time. In this example, memristive hysteresis is caused by the redistribution of mobile vacancies. **b** Side view of the memristive devices considered in this work. **c** Simulated current $I(t)$ together with the applied voltage $V(t)$ over two voltage cycles, using $MoS_2$ as the channel material. **d** Electrostatic potential and vacancy density profiles at three selected points (I-III) during the second cycle, demonstrating the delayed response of vacancies, the formation of a depletion region at the right contact, and accumulation at the left contact. White arrows indicate the direction of the negative potential gradient. **e** Corresponding current-voltage characteristic for the second cycle, with points I-III indicated.

**Parameterization and dataset generation.** The charge transport model provides a systematic parameterization of the memristive system through its material, geometric, interfaces, and operating parameters. These parameters serve as input variables for the simulations and define the physical state and external conditions of the device. Parameters related to microstructure and interfaces, such as mobilities and Schottky barriers, are particularly important, as they can vary, often over orders of magnitude, depending on fabrication processes [32–35]. Many of these quantities are challenging to determine experimentally, making them valuable targets for inverse modeling and model-guided device optimization. We vary nine key parameters: electron mobility $\mu_\mathrm{n}$, vacancy mobility $\mu_\mathrm{x}$, Schottky barriers $\phi_{0,\mathrm{L}}$ and $\phi_{0,\mathrm{R}}$, background donor density $n_\mathrm{d}$, vacancy energy offset $\delta_\mathrm{x}$ relative to the conduction band edge, channel length $L$, and two voltage parameters describing the triangular voltage sweep $V(t)$, namely the period $t_\mathrm{p}$ and maximum voltage $V_\mathrm{max}$ (Figure 2a, inset). The vacancy energy offset $\delta_\mathrm{x}$ controls the average sulfur vacancy density $n_\mathrm{x}$. High-throughput I-V curve simulations are performed by pseudo-randomly varying these parameters across physically motivated ranges based on experimental and theoretical studies of layered TMDCs [32–37]. Full 9D datasets and lower-dimensional subsets (7D, 5D, 4D, 2D) are generated, each comprising at least 10,000 simulations. A summary and further details are provided in the Methods.

**ANN surrogate model.** The datasets generated by the finite-volume simulations are used to train a sequence-to-sequence artificial neural network (ANN) that predicts the dynamic current response from the voltage and static device parameters, as illustrated in Figure 3. The model consists of one or two stacked LSTM layers [61], followed by dropout regularization and a fully connected output layer. Hidden layer sizes vary from 64 to 196 units depending on the dataset.

The network receives the scaled voltage sequence $V_{\mathrm{n}}$, obtained from $V(t)$, together with a set of scaled, time-independent input parameters $x_{\mathrm{n}}$ with $x \in \{L, n_{\mathrm{d}}, \delta_{\mathrm{x}}, \mu_{\mathrm{n}}, \mu_{\mathrm{x}}, \phi_{0,\mathrm{L}}, \phi_{0,\mathrm{R}}\}$. Both the voltage sequence and the scalar parameters are scaled to the interval [0,1] using min-max scaling based on the dataset ranges. Scalar inputs are expanded across the time dimension to form a multichannel input sequence. Since the voltage sequence encodes two physical quantities (period and maximum voltage) but enters as a single input channel, the total number of input channels is one less than the dataset dimensionality.

The output is the scaled current sequence $I_{\mathrm{n}}$, obtained by applying a custom scaling procedure to the original current response $I(t)$. Because the current magnitudes of different sequences vary by orders of magnitude, a suitable scaling transformation is essential for robust training. However, standard log-normal standardization is not readily applicable due to zero crossings in the current sequences. Instead, we scale sequences based on a fitted log-normal distribution of current maxima across the dataset. This approach preserves the overall shape and polarity of the I-V curves, ensures balanced error weighting during training, and avoids distortions associated with alternative offset-based methods.

Each dataset is divided into training, validation, and test subsets, with the test set used exclusively to evaluate model performance on previously unseen parameter combinations. Models are trained using the Adam optimizer [62] with mean squared error loss, and hyperparameters selected by randomized search. Performance is evaluated using the normalized mean absolute error (nMAE), providing a relative error metric constrained to [0,1] for comparability across different current regimes. Further details are provided in the Methods section.

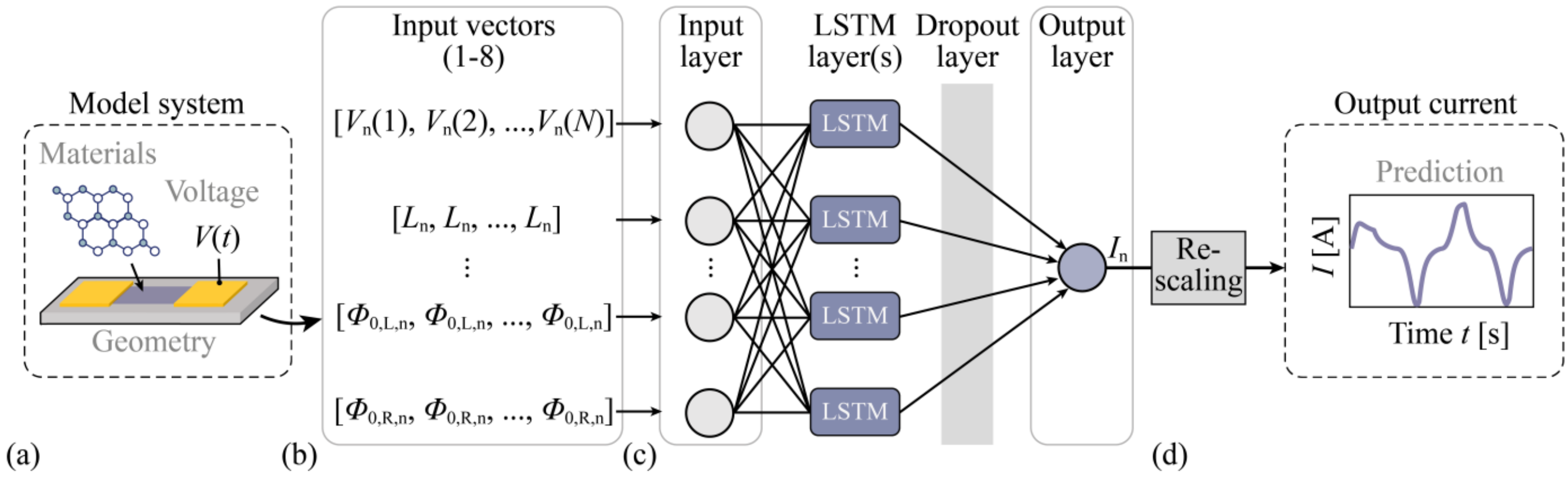

**Figure 3. Illustration of the model workflow and the ANN architecture. a** The device is parameterized with a finite-volume charge transport model, which is used to generate datasets with variations in the material, device, and voltage parameters. **b** The time dependent voltage vectors are scaled, resampled to a length $N$, and used as the first input feature vector. The up to seven other time-independent input features are scaled and extended in the time domain to match the length $N$ of the scaled voltage vector. **c** An LSTM-based ANN is trained using the scaled vectors as input and the scaled current $I_n$ as a target vector. **d** The predicted $I_n$ is rescaled to obtain the current $I(t)$ as a function of the time $t$. Because the voltage $V(t)$ is known as an input feature, I-V curves can be obtained from $I(t)$.

### 3.2 Performance Evaluation

Depending on the intended application, datasets of varying dimensionality can be used. For example, in real-time circuit simulations or large-scale array modeling involving thousands of individual or coupled devices, the voltage is a critical input parameter, while other inputs not required for the respective optimization task may be omitted. Conversely, in parameter fitting tasks, the voltage and device dimensions are often known and thus can be excluded from the input feature set. In general, lower-dimensional datasets are expected to require fewer training samples and less training effort, albeit at the cost of reduced model generality. Understanding model performance for different input dimensionalities and identifying the required data for accurate prediction are therefore essential for evaluating the feasibility of different application scenarios.

To assess model performance across different input dimensions, we train the ANN on the datasets with 2, 4, 5, 7, and 9 input features, respectively. Each dataset is split into 80% training, 10% validation, and 10% testing subsets, with a total of 10,000 samples for the 2D and 4D datasets and 20,000 samples for the 5D, 7D, and 9D datasets. Model performance is evaluated using the normalized mean absolute error (nMAE), reported as a percentage (%). The resulting test nMAEs of the best-performing training trials are shown in Figure 4a, where predictions are sorted in ascending order by nMAE. In each case, the error distribution exhibits a narrow low-error region and a broader high-error tail comprising approximately the first and last 5-10% of test samples. Across all datasets, nMAEs range from approximately 0.3% (minimum, 7D dataset) to 3.4% (maximum, 9D dataset), with typical values $< 1\%$ and qualitatively similar error distribution trends. The mean and median nMAEs are summarized in Figures 4b. A clear trend emerges: the mean nMAE increases with input dimensionality, rising from approximately 0.4% (2D-5D) to 0.9% (9D). Similarly, the median nMAE is lowest for the 2D-5D datasets at approximately 0.3%, and increases by a factor of two to three for the 7D and 9D datasets.

To illustrate prediction quality, representative examples spanning the nMAE range of the 5D model are shown in Figures 4c-f, including various voltage cycles and degrees of nonlinearity. For samples with small and intermediate nMAEs (ranging from 0.2% to 0.8%; Figures 4c-e), the predicted I-V curves closely match the ground truth from the charge transport model. Even for the largest errors in the test set, deviations remain relatively minor and are mostly confined to the first voltage cycle Figures 4f. This slightly larger initial error likely reflects the increased nonlinearity of the current response during the first voltage cycle, caused by the stronger redistribution of charge carriers from their initial equilibrium configuration before the system reaches a near-steady-state oscillation [42]. Overall, the results confirm that the model captures the nonlinear I-V behavior reliably and with high fidelity across a broad range of conditions.

**Inference time.** In addition to prediction accuracy, inference time is a critical factor for applying the ANN model to high-throughput tasks such as inverse modeling, parameter space exploration, and design optimization. To evaluate performance, we conduct 10,000 randomized predictions across the parameter space using an ANN trained on the 5D dataset and measure the prediction time. In parallel, we simulate 1,000 parameter combinations using the full charge transport model. The ANN achieves an average inference time of $0.0013$ s $\pm$ $0.0002$ s, whereas the finite volume model requires $30.2$ s $\pm$ $0.7$ s per simulation, both evaluated on a single CPU system. This corresponds to an approximate speedup of 20,000×, equivalent to a gain of more than four orders of magnitude compared to the original charge transport model. Such a substantial improvement enables application scenarios that would be computationally infeasible with the full numerical model. Several of these use cases are explored in the following sections.

Further acceleration can be achieved by exploiting the batch processing capabilities of standard AI frameworks, which enable parallel evaluation of large numbers of input samples on GPUs. Using batch sizes of 10,000, the average inference time per I-V curve is reduced to approximately 1.9 microseconds, corresponding to a total acceleration of over seven orders of magnitude relative to the finite volume model. GPU-based batch inference is especially beneficial for highly parallelizable tasks such as parameter mapping and global sensitivity analysis. For consistency, all studies presented in Sections 3.2-3.4 were conducted using CPU-based inference.

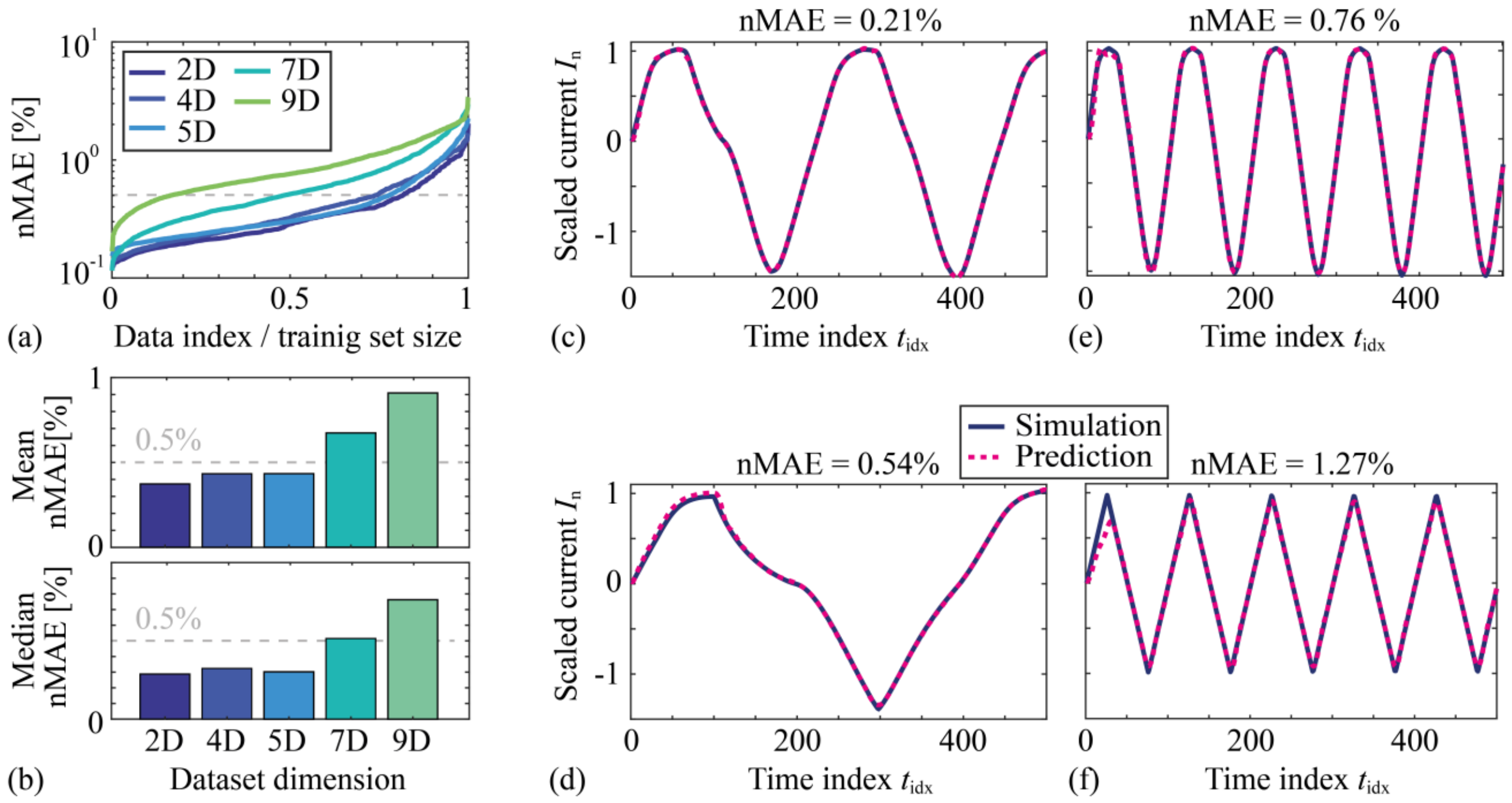

**Figure 4. ANN model performance across datasets of increasing input dimensionality.** **a** Normalized mean absolute error (nMAE) between predicted and true test current sequences, sorted in ascending order across the five datasets. The free parameters and their respective ranges are summarized in the Methods section. **b** Mean and median test nMAEs corresponding to the results in (a) for the five datasets with 2, 4, 5, 7, and 9 input features, respectively. **c-f** Comparison between example ANN predictions (dashed lines) and ground truth finite volume simulations (solid lines) for representative test samples at different nMAE levels, ranging from 0.21% to 1.27%.

### 3.3 Application Example I: Validation & Inverse Modeling

As a first application example and validation of the ANN model, we consider an inverse optimization task in which model parameters are extracted by fitting simulations to experimental I-V characteristics of two-terminal lateral memristive devices based on $MoS_2$, as reported in Ref. [59]. This type of inverse problem is commonly used for model validation and device analysis, as it enables the connection of physical semiconductor parameters with material design, device geometry, and fabrication conditions. In this context, the device geometry and operating conditions, such as channel length, maximum applied voltage, and voltage period, are typically controlled during fabrication and measurement. These quantities are therefore treated as known, reducing the dimensionality of the optimization problem from nine to five. The remaining parameters, electron mobility, the two Schottky barrier heights, vacancy mobility, and the intrinsic vacancy energy level, are strongly influenced by fabrication-dependent variations in material composition, microstructure, and interface properties [32–35]. These quantities are also challenging to access experimentally, in part because hysteresis effects prevent the use of standard analytical methods [63,64].

To solve this inverse problem, we apply the trained ANN model as a fast, differentiable surrogate for the underlying charge transport model. We use a global optimization method that combines randomized starting points with a nonlinear solver for local refinement [65], which helps identify suitable parameter sets despite the nonlinearity of the ANN model. Using the ANN surrogate model reduces the computational cost per evaluation by several orders of magnitude compared to the finite volume solver. Since each optimization run typically requires thousands to tens of thousands of evaluations, this speedup enables parameter extraction to be completed within seconds or minutes rather than weeks or months. For example, the inverse modeling tasks described below each required approximately 75,000 surrogate model evaluations and were completed in about 110 seconds, compared to an estimated 26 days using the finite volume model.

Results from two inverse modeling procedures, based on measurements from distinct devices reproduced from Li et al. [59], are shown in Figure 5a,b. The fitted model parameters corresponding to these devices are summarized in Table 4 (Methods). In each case, the fitted ANN model captures the overall hysteresis shape and current amplitude, with deviations primarily observed on the positive voltage branch. The nMAEs between prediction and experiment are approximately 3.4% (Figure 5a) and 6.3% (Figure 5b), which is well within the expected variation of the experimental I-V curves [66–68]. As a final verification, we input the fitted parameter sets into the original finite volume charge transport model. The resulting simulations closely match the ANN predictions, confirming that the accuracy of the ANN surrogate is sufficient for inverse modeling tasks and that the remaining discrepancies arise from model limitations or experimental variability, not from the surrogate itself. Combined with the significant acceleration compared to finite volume simulations, this accuracy enables practical high-throughput parameter extraction across large experimental datasets.

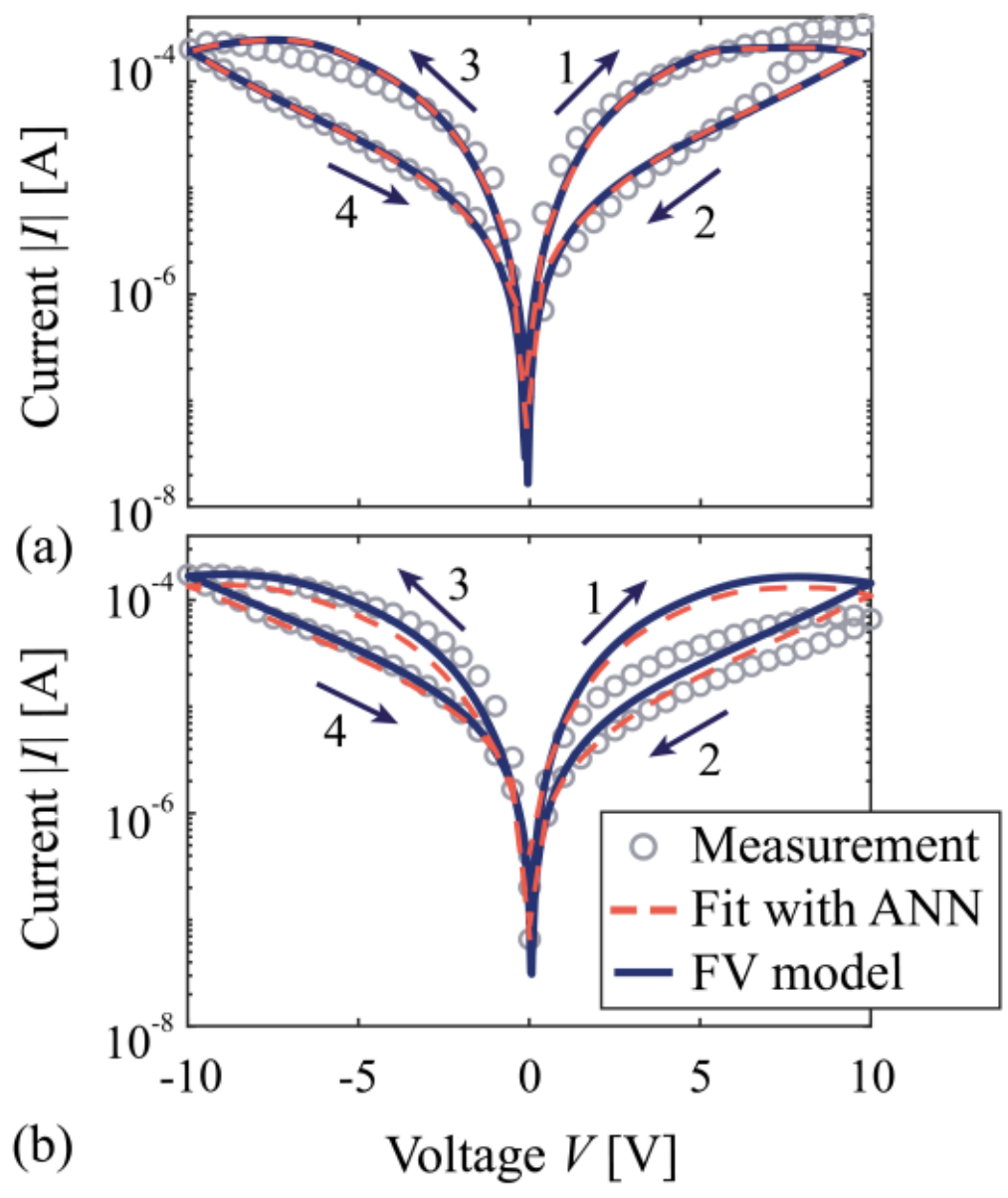

**Figure 5. Inverse modeling of experimental I-V curves using the ANN surrogate model.** **a, b** Measured current-voltage characteristics (circles) from two lateral $MoS_2$-based memristive devices, reproduced from Li et al. [59], fitted using the ANN surrogate model (red, dashed) by optimizing five physical model parameters. The fitted parameter sets are subsequently evaluated with the finite volume charge transport model (blue) for verification. The normalized MAEs between measurement and ANN prediction are 3.4% (a) and 6.3% (b), which is well within the expected variation of the experimental I-V curves[66–68]. The extracted model parameters are listed in Table 4 (Methods).

### 3.4 Application Example II: Parameter Space Exploration

As a second application example, we use the model to explore the parameter space and analyze how the input parameters influence application-relevant device metrics. These metrics are typically scalar quantities used to compare different devices and assess their suitability for specific use cases. For I-V curves, common examples include symmetry, on-off ratio, and linearity, each intended to capture a specific aspect of device behavior [27,69–71]. For instance, symmetry can serve as a proxy for rectifying behavior, which is relevant in crossbar arrays to suppress sneak path currents [69]. Likewise, the on-off ratio is an established metric for quantifying the extent of hysteresis by comparing the current magnitude at a fixed voltage in the high- and low-conductance branches [70,71]. To illustrate how the trained ANN model enables systematic parameter space exploration, we present two complementary approaches using symmetry and on–off ratio as example metrics: mapping their variation across selected input dimensions and performing a variance-based global sensitivity analysis.

**Symmetry maps.** To quantify the symmetry $S$ of the I-V curves, we define it as the absolute difference in the normalized current magnitudes at maximum and minimum voltage. This yields a range $0 \leq S \leq 1$, with $S = 1$ corresponding to perfect symmetry, i.e., equal current magnitudes under positive and negative maximum bias. To explore the symmetry landscape, we use the trained ANN model to perform predictions over 25,000 input conditions, requiring approximately 33 seconds, compared to an estimated 8.7 days using the finite volume charge transport model. The input conditions were generated by varying the Schottky barriers $\phi_{0,\mathrm{L}}$, $\phi_{0,\mathrm{R}}$, and the vacancy mobility $\mu_{\mathrm{x}}$. A selection of the resulting symmetry maps $S(\phi_{0,\mathrm{L}}, \phi_{0,\mathrm{R}})$ is shown in Figure 6a-f for six different values of $\mu_{\mathrm{x}}$.

At low vacancy mobilities (Figure 6a–c), the symmetry maps exhibit a well-defined high-symmetry band along the diagonal $\phi_{0,\mathrm{L}} \approx \phi_{0,\mathrm{R}}$, ranging from 0.05 eV to 0.15 eV. This indicates that symmetric contact barriers yield symmetric I-V responses when ionic motion is limited. As $\mu_{\mathrm{x}}$ increases (Figure 6d–f), the high-symmetry region broadens (initially in the low-barrier region) and becomes increasingly diffuse, eventually covering nearly the entire parameter space at $\mu_{\mathrm{x}} = 1 \times 10^{-13}\ \mathrm{m^2/(Vs)}$ (Figure 6f). To validate the symmetry metric, selected I-V curves corresponding to three representative off-diagonal parameter points are shown in Figure 6g–i. These points, marked in the symmetry maps (Figure 6d–f), correspond to increasing mobilities of approximately $\mu_{\mathrm{x}} \approx 3.6 \times 10^{-14}$, $6 \times 10^{-14}$, and $1 \times 10^{-13}\ \mathrm{m^2/(Vs)}$, respectively. Although the overall shape of the I-V curves remains qualitatively similar, the degree of symmetry increases with $\mu_{\mathrm{x}}$, consistent with the trend observed in the maps.

The observed broadening of the symmetry band with increasing vacancy mobility can be understood by considering the dominant current-limiting mechanism. At high Schottky barriers, the current is primarily controlled by the contacts, and symmetry is governed by the relative barrier heights. At low Schottky barriers, channel transport becomes dominant, particularly the contribution of the positively charged mobile vacancies. As $\mu_x$ increases, the applied field induces a stronger redistribution of vacancies, leading to dynamic depletion zones that limit the current asymmetrically. This effect suppresses the influence of the contact asymmetry, resulting in more symmetric I-V curves even for asymmetric Schottky barriers. This interpretation aligns with previous observations of vacancy-dynamics-induced changes in hysteresis area [42].

**Sensitivity analysis: symmetry.** While these parameter maps provide useful insight into the role of vacancy mobility and contact asymmetry, they cover only a small subset of the full input space. Such an analysis is inherently constrained to low-dimensional projections and offers limited scope for systematically quantifying each parameter's contribution to the symmetry metric. To extend the analysis to the full input space, we performed a variance-based global sensitivity analysis using Sobol' indices [72–74]. This approach is well suited to memristive devices, where complex, hysteretic transport processes lead to nonlinear parameter dependencies that cannot be captured accurately by linear correlation.

Sobol' indices decompose the variance of the metric into contributions from individual parameters and their combinations: the total index captures the overall contribution of a parameter, the first-order index quantifies its direct effect, and higher-order indices resolve joint contributions from multiple parameter combinations. Including higher-order indices enables identifying non-additive dependencies in the model response but substantially increases computational cost due to the need for high-dimensional sampling, typically requiring tens to hundreds of thousands of model evaluations for many input parameters [73,75].

We use the trained ANN model to evaluate the symmetry metric across the full parameter space, including the total, first-order, and selected second-order Sobol' indices, as shown in Figure 6j-l. The analysis is based on 175,000 model evaluations across the five-dimensional parameter space spanned by $\delta_{\mathrm{x}}$, $\mu_{\mathrm{n}}$, $\mu_{\mathrm{x}}$, $\phi_{0,\mathrm{L}}$, and $\phi_{0,\mathrm{R}}$, and was completed in 5.6 minutes using the ANN surrogate model, compared to an estimated 59 days with the finite volume charge transport model (see Methods).

The analysis of the total sensitivity indices reveals that three of the five parameters contribute significantly to the variation in the symmetry metric. The Schottky barrier heights $\phi_{0,\mathrm{L}}$ and $\phi_{0,\mathrm{R}}$ each account for approximately 50% of the total output variance, while the electron mobility $\mu_{\mathrm{n}}$ exhibits a slightly higher total contribution of approximately 62%. In contrast, the vacancy mobility $\mu_{\mathrm{x}}$ contributes only about 17%, and the vacancy energy offset $\delta_{\mathrm{x}}$ has a negligible effect. Since the total sensitivity indices include contributions from all interaction effects, their sum exceeds 100%, reflecting the presence of strong higher-order interactions.

The analysis of the first-order sensitivity indices shows that the direct contributions of all parameters are considerably smaller than their total effects. The first-order index of $\mu_{\mathrm{n}}$ is approximately 37%, while those of the Schottky barriers and the vacancy mobility are insignificant, below 5%. This indicates that interactions between parameters play a major role in determining the symmetry behavior. Analysis of the second-order indices further reveals a strong joint contribution between $\phi_{0,\mathrm{L}}$ and $\phi_{0,\mathrm{R}}$ ($\approx$ 21%), whereas second-order interactions involving $\mu_{\mathrm{n}}$ or $\mu_{\mathrm{x}}$ remain small. This suggests that the influence of $\mu_{\mathrm{n}}$ arises predominantly through higher-order interactions beyond second order. In contrast, the dominant contributions of the Schottky barriers are associated with both second-order interactions between $\phi_{0,\mathrm{L}}$ and $\phi_{0,\mathrm{R}}$ and higher-order effects, while the vacancy mobility $\mu_\mathrm{x}$ has a small effect and the vacancy energy offset $\delta_\mathrm{x}$ has negligible impact on symmetry overall.

Hence, consistent with the symmetry maps, the Schottky barriers $\phi_{0,\mathrm{L}}$ and $\phi_{0,\mathrm{R}}$ are the dominant factors controlling symmetry. The vacancy mobility $\mu_\mathrm{x}$ shows a smaller contribution, leading to a broadening of the high-symmetry region at higher values. The electron mobility $\mu_\mathrm{n}$ also contributes according to the sensitivity analysis, likely reflecting limitations in electronic transport at low mobility. Overall, symmetry is primarily controlled by interface transport and secondarily influenced by electronic transport in the channel.

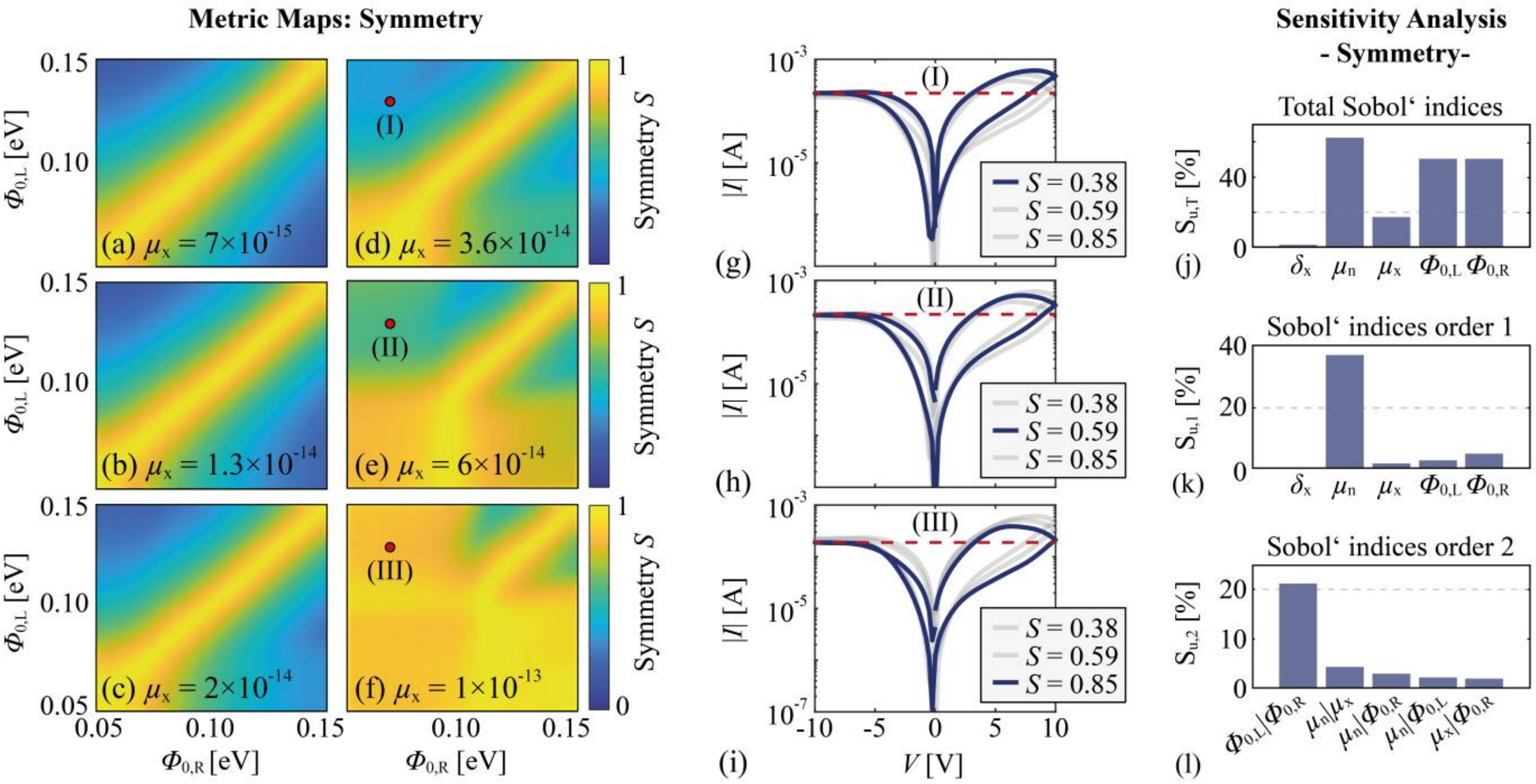

**Figure 6. Application of the physics-guided ANN model to evaluate the symmetry metric and its parameter sensitivities. a-f** Selected symmetry maps $S(\phi_{0,\mathrm{L}}, \phi_{0,\mathrm{R}})$ at six different values of the vacancy mobilities $\mu_\mathrm{x}$, obtained from a high-throughput parameter sweep using the trained ANN model across approximately 25,000 combinations of $\phi_{0,\mathrm{L}}, \phi_{0,\mathrm{R}}$, and $\mu_\mathrm{x}$. At low $\mu_\mathrm{x}$, a narrow high-symmetry band appears along the diagonal $\phi_{0,\mathrm{L}} \approx \phi_{0,\mathrm{R}}$, which

broadens progressively with increasing mobility. All values for $\mu_x$ are provided in units of $m^2/(Vs)$. **g-i** Example I-V curves at selected off-diagonal parameter points (highlighted in panels d-f) for three representative mobilities, illustrating how increasing $\mu_x$ leads to more symmetric current responses despite asymmetric Schottky barriers. **j-l** Sobol' sensitivity analysis of the symmetry metric over the 5D parameter space using 175,000 ANN evaluations, showing total Sobol indices $S_{u,T}$, first-order indices $S_{u,1}$, and the five largest second-order indices $S_{u,2}$. Parameters include electron mobility $\mu_n$, vacancy mobility $\mu_x$, Schottky barriers $\phi_{0,L}, \phi_{0,R}$, and vacancy energy offset $\delta_x$.

**On-off ratio maps.** As a second example metric, we define a normalized on-off ratio $R$ to quantify the extent and direction of hysteresis at a fixed voltage of $V = 5$ V in the right branch of the hysteresis loop. The metric is constructed such that clockwise hysteresis corresponds to the interval $R \in [0.5, 1]$, with $R = 1$ indicating an ideal open loop with maximum hysteresis and $R = 0.5$ indicating no hysteresis. For $R < 0.5$, the direction of the hysteresis is reversed (counterclockwise), with $R = 0$ representing an ideally inverted loop with maximum hysteresis. This formulation offers two key advantages for optimization and sensitivity analysis: it is strictly bounded between 0 and 1 and allows for direct discrimination of the hysteresis direction.

We extract $R$ from the same ANN simulations used for the symmetry analysis. Selected metric maps $R(\phi_{0,L}, \phi_{0,R})$ at six different vacancy mobilities $\mu_x$ are shown in Figure 7a-f. At low vacancy mobilities up to $\mu_x \approx 10^{-14}$ $m^2/(Vs)$, the maps are largely featureless, with $R \approx 0.5$ across the entire Schottky barrier space, indicating an absence of significant hysteresis (Figure 7a,b). As $\mu_x$ increases to several $10^{-14}$ $m^2/(Vs)$, a clear structure begins to emerge: regions of enhanced hysteresis form in the lower half of the maps, particularly at small $\phi_{0,L}$, and gradually expand across the barrier space (Figure 7c-f). This development reflects the increasing onset of field-driven vacancy migration, which becomes more pronounced as mobility increases and enables stronger modulation of the channel conductivity during the sweep. The increasing influence of $\phi_{0,\mathrm{L}}$ indicates that the Schottky barrier height influences the restoring force acting on the redistributed vacancies, thereby modulating their dynamic response that causes hysteresis [42]. Its larger influence on the hysteresis compared to $\phi_{0,\mathrm{R}}$ reflects the fact that each hysteresis branch is predominantly shaped by one contact barrier, and our definition of the on-off ratio focuses on the right branch ($V \geq 0$ V), where $\phi_{0,\mathrm{L}}$ governs the space charge distribution and modulates the driving and restoring force acting on the vacancies during the sweep. Example I-V curves from selected points in the map at $\mu_{\mathrm{x}} \approx 6 \times 10^{-14}\ \mathrm{m}^2/(\mathrm{Vs})$ (Figure 7e) are shown in Figure 7g-i. At $R = 0.58$, the hysteresis is barely visible, while curves with $R = 0.76$ and $R = 0.88$ exhibit clear, pronounced hysteresis. These results confirm that our definition of $R$ captures the emergence of hysteresis in a physically meaningful way.

**Sensitivity analysis: on-off ratio.** To quantify the parameter dependencies of the on-off ratio, we perform a second Sobol' sensitivity analysis across the same parameter space and number of evaluations as before. The resulting total, first-order, and second-order Sobol' indices are shown in Figure 7j-l. The total Sobol' indices (Figure 7j) show that the vacancy mobility $\mu_{\mathrm{x}}$ dominates the variation in the on-off ratio, with a total contribution of 87%. This trend is consistent with the observed evolution in the metric maps (Figure 7a–f), where increasing $\mu_{\mathrm{x}}$ leads to a pronounced transition from nearly uniform behavior to pronounced variation in the on-off ratio across the Schottky barrier space. The first-order Sobol' index of $\mu_{\mathrm{x}}$ (68%) confirms that its influence is primarily direct. Interaction terms involving $\mu_{\mathrm{x}}$, particularly with $\mu_{\mathrm{n}}$ and $\phi_{0,\mathrm{L}}$, contribute an additional $\approx 10\%$, explaining the gap between total and first-order sensitivity. The electron mobility $\mu_n$ and the left Schottky barrier $\phi_{0,L}$ show moderate total effects ($\approx 20\%$), while all other parameters, including the vacancy energy offset $\delta_x$ and the right Schottky barrier $\phi_{0,R}$, have negligible influence. The stronger role of $\phi_{0,L}$ compared to $\phi_{0,R}$ is also visible in the metric maps and reflected in the Sobol' indices (21% vs. 3%).

Taken together, the results indicate that the on-off ratio is primarily determined by vacancy dynamics, with smaller contributions from carrier mobility and contact asymmetry. While image-charge-induced Schottky barrier lowering is included, variations in barrier height do not dominate the hysteresis behavior, as quantified by the on-off ratio, across the parameter space. Instead, vacancy mobility controls the dynamic modulation of channel conductivity. This finding is consistent with previous indications that hysteresis arises from vacancy depletion and redistribution dynamics in the channel [42], and is here quantitatively confirmed across a broad parameter range using the ANN model.

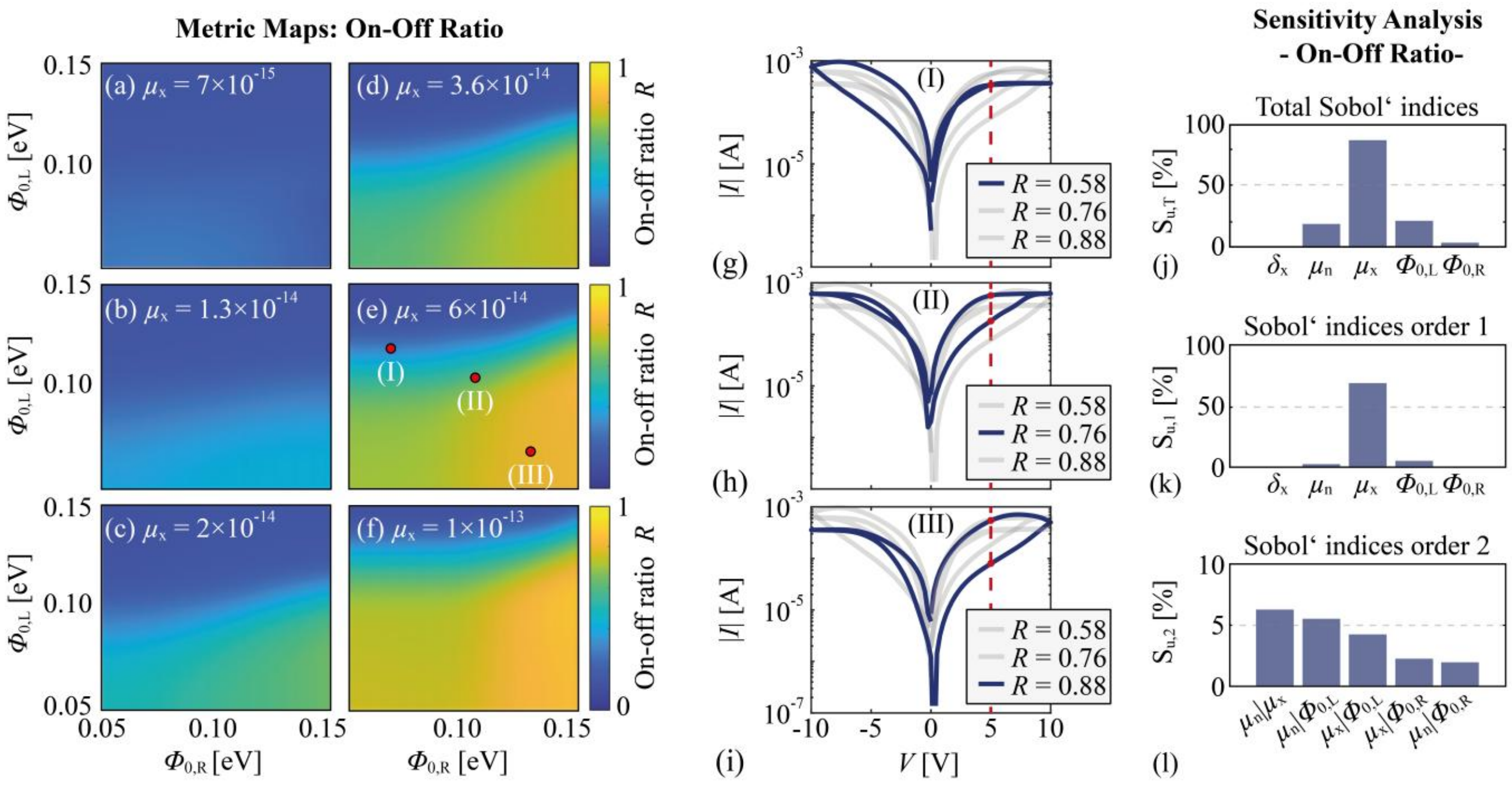

**Figure 7. Application of the physics-guided ANN model to evaluate the on-off ratio metric and its parameter sensitivities. a-f** Normalized on-off ratio maps $R(\phi_{0,L}, \phi_{0,R})$ at six

different values of the vacancy mobility $\mu_x$, obtained from high-throughput ANN evaluations over 25,000 combinations of $\phi_{0,L}$, $\phi_{0,R}$, and $\mu_x$. At low mobilities, $R \approx 0.5$ across the space, indicating negligible hysteresis. As $\mu_x$ increases, structured regions of enhanced hysteresis emerge, especially at small $\phi_{0,L}$. All values for $\mu_x$ are in units of m²⁄(Vs). **g–i** Representative I-V curves at three off-diagonal points in the map at $\mu_x \approx 6 \times 10^{-14}\ \mathrm{m^2/(Vs)}$ illustrating the increasing hysteresis magnitude with increasing $R$. **j–l** Sobol' sensitivity analysis of the on-off ratio metric over the 5D parameter space using 175,000 ANN evaluations, showing total Sobol indices $S_{u,T}$, first-order indices $S_{u,1}$, and the five largest second-order indices $S_{u,2}$. The results confirm that the on-off ratio is governed primarily by $\mu_x$, with small contributions from $\mu_n$ and $\phi_{0,L}$, and negligible influence from other parameters.

### 3.5 Application Example III: Design Optimization

In the previous section, we introduced the symmetry metric $S$ and the on-off ratio $R$ as example performance metrics derived from the simulated I-V characteristics. We showed that each captures relevant features of the device behavior and analyzed how they depend on the underlying design parameters. Depending on the intended application of the memristive device, different combinations of these metric values may be desirable. For instance, neuromorphic and memory systems typically favor a large on-off ratio, while strong asymmetry can be advantageous to mitigate sneak-path currents in crossbar arrays [69,71].

The global sensitivity analysis demonstrated, using $S$ and $R$ as examples, that different metrics are primarily influenced by different combinations of design parameters and their interactions. Despite these differences, all these metrics are ultimately derived from the same I-V characteristics and governed by the same underlying charge transport physics. As a result, they are not fully independent, and identifying parameter configurations that achieve specific combinations of metrics becomes a nontrivial task that can involve trade-offs. This task becomes increasingly challenging as the number of free parameters and design objectives grows.

To examine how such multi-metric design challenges can be addressed using the proposed framework, we apply the physics-guided ANN model to a representative case involving the symmetry metric $S$ and the on-off ratio $R$. We first define a scalarized single-objective cost function that combines the two metrics, with the ideal solution corresponding to $S = 0$ (maximum asymmetry) and $R = 1$ (maximum on-off ratio). The resulting cost landscape is shown in Figure 8a. Using the global optimization approach described in Section 3.2, we obtain a solution (red dot) after approximately 77,000 surrogate model evaluations, completed in 145 seconds, compared to an estimated 27 days using the charge transport model. The corresponding I-V curve and ANN-predicted parameters are shown in Figure 8b. As intended, the result exhibits a large on-off ratio ($R \approx 0.87$) and small symmetry ($S \approx 0.15$), consistent with the design objective. Notably, the identified optimum does not lie at the global minimum of the cost function, which reflects an inherent trade-off between symmetry and on-off ratio due to their shared physical dependencies.

To confirm and further explore the trade-off between $S$ and $R$, we formulate a multi-objective optimization problem treating both metrics as independent objectives. The Pareto front is computed using a population-based evolutionary algorithm [76], resulting in approximately 40,000 surrogate model evaluations, completed in 73 seconds, compared to an estimated 14 days using the finite volume charge transport model. As shown in Figure 8a (white points), the Pareto front intersects the single-objective solution, confirming its Pareto-optimality. The front reveals a clear trade-off: achieving lower symmetry (greater asymmetry) requires a reduction in on-off ratio, and vice versa. This example demonstrates how the ANN surrogate enables rapid, interpretable design optimization. The approach is readily extensible to include additional performance metrics or physical constraints, supporting more complex inverse design tasks.

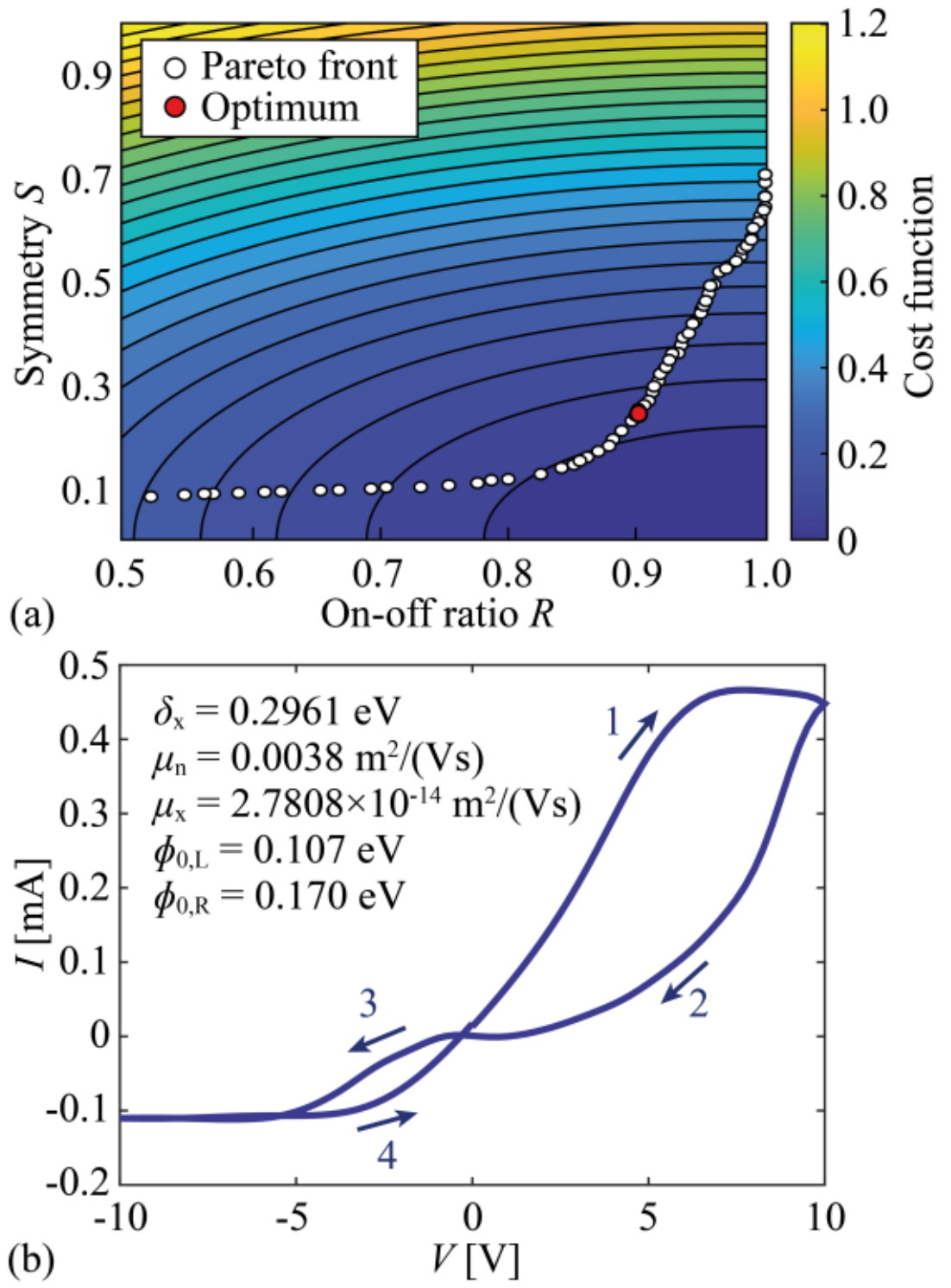

**Figure 8. Optimization and trade-off analysis enabled by the physics-guided ANN model**. **a** Cost function map for a scalarized single-objective optimization, combining the symmetry metric $S$ and the on-off ratio metric $R$ of the right hysteresis branch into a single objective to minimize $S$ and maximize $R$. The Pareto front (white points), obtained from a separate multi-objective optimization with 40,000 ANN evaluations within 73 s, illustrates the trade-off between the two metrics. The red dot marks the optimum identified by the scalarized optimization, which lies on the Pareto front and reflects a specific compromise between symmetry and on-off ratio. **b** Simulated I-V characteristics at the selected optimum in (a),

along with the corresponding physical model parameters and arrows indicating the hysteresis direction.

## 3 DISCUSSION

We presented a physics-guided modeling framework that combines high-fidelity finite-volume (FV) charge transport simulations with LSTM-based artificial neural networks (ANNs) to predict dynamic current-voltage behavior in 2D memristive devices. Trained on physically grounded data describing vacancy-dominated switching in lateral $MoS_2$ structures, the ANN surrogate reduces simulation time from approximately 30 seconds to 1 millisecond per I-V curve, while maintaining physical interpretability and high predictive accuracy, with mean and median normalized mean absolute errors (nMAEs) < 1%. A custom scaling procedure enables consistent training across sequences with large current variations and zero crossings, preserving sequence structure and balanced error weighting.

Three key applications demonstrated the versatility of the framework: (i) inverse modeling for parameter extraction and validation, (ii) design space exploration via metric mapping and sensitivity analysis, and (iii) multi-objective performance optimization. By enabling these applications, the framework provides systematic access to physical insights linking device behavior to underlying transport mechanisms. Inverse modeling reproduced experimental I-V characteristics with good agreement. Design space mapping and global sensitivity analysis identified symmetry as primarily governed by Schottky barrier interactions and electron mobility, while the on-off ratio is dominated by vacancy mobility. Multi-objective optimization revealed explicit trade-offs between symmetry and on-off ratio, illustrating model-driven design optimization and supporting more advanced, application-specific design scenarios.

Across all tasks, involving tens to hundreds of thousands of model evaluations, the ANN surrogate enabled completion within seconds to minutes on CPUs, compared to days to months with finite-volume simulations. GPU-based batch inference further accelerated large-scale evaluations to less than a second for hundreds of thousands of samples, achieving speedups exceeding seven orders of magnitude. A summary is provided in Table 1.

Together, these results demonstrate that the physics-guided ANN enables a shift from simulation-limited to exploration-driven modeling workflows, allowing full optimization and analysis tasks to be completed in seconds or minutes instead of weeks or months. The framework provides an efficient, scalable, and physically interpretable extension to high-fidelity simulation workflows for emerging memristive devices.

While this study focuses on lateral $MoS_2$-based structures, the approach is broadly extensible to alternative materials [21,77,78], multi-terminal configurations [17,23,79,80], and gate-tunable synaptic memtransistors [22,81,82]. The scaling procedure may also support stable training in other dynamic modeling tasks involving sign-changing quantities. In combination with experimental measurements, the framework may support closed-loop workflows for accelerated device optimization and high-throughput screening across diverse material and device systems. Extension to other classes of hysteretic electronic devices would require adapting the underlying charge transport model to capture material-specific mechanisms, while retaining the sequence modeling framework. Beyond the specific demonstration on 2D memristive devices, the framework establishes a generalizable approach for rapid, interpretable modeling of dynamic, nonlinear, and hysteretic electronic systems.

**Table 1.** Comparison of ANN inference times and finite volume (FV) simulation times for five representative tasks: (1) simulating a single I-V curve, (2) parameter extraction via inverse modeling, (3) parameter space mapping (metric maps), (4) global sensitivity analysis, and (5) constrained design optimization. ANN inference was performed either on a CPU (Intel® Core™ i9-14900KF) or in parallel on a GPU (NVIDIA® GeForce™ RTX 4090, batch size 10,000). Details are provided in Sections 3.2-3.5 and Methods.

| | **Prediction** | **Parameter extraction** | **Parameter space exploration** | | **Inverse design optimization** | |
|---|---|---|---|---|---|---|
| | Single I-V curve | Single-objective | Metric maps | Sensitivity analysis | Single-objective | Pareto optimization |
| ANN (GPU) | 1.9 μs ± 8ns | - | 0.05 s | 0.32 s | - | - |
| ANN (CPU) | 1.3 ± 0.2 ms | 110 s | 33 s | 5.6 min | 145 s | 73 s |
| FV (CPU) | 30.2 ± 0.7 s | 26 days | 8.7 days | 59 days | 27 days | 14 days |
| No. evaluations | 1 | 75,000 | 25,000 | 170,000 | 77,000 | 40,000 |

# 4 METHODS

## 4.1 Charge transport model

The model physics is described by a system of coupled partial differential equations, which are discretized via the finite volume method in the spatial domain and implemented in the Julia programming language utilizing the VoronoiFVM.jl solver [83–85]. The finite volume method is particularly well suited for these conservative flow problems, as it ensures local flux conservation, thermodynamic consistency, and avoids unphysical negative densities [57,58]. Specifically, we solve the three drift-diffusion equations for electrons n, holes p and vacancies x

$$q_\alpha \partial_{\mathrm{t}} n_\alpha + \nabla \cdot \boldsymbol{j}_\alpha = 0 \quad \text{with} \quad \alpha \in \{\mathrm{n}, \mathrm{p}, \mathrm{x}\}, \tag{1}$$

with the corresponding charge carrier densities $n_\alpha$, the partial derivative operator $\partial_{\mathrm{t}} \coloneqq \partial/\partial t$ with respect to the time $t$, and their respective charges $q_\alpha = q z_\alpha$ with the charge numbers $z_\alpha$ ($z_{\mathrm{n}} = -1$, $z_{\mathrm{p}} = z_{\mathrm{x}} = +1$) and the positive elementary charge $q$. The current densities $\boldsymbol{j}_\alpha$ are expressed via the quasi-Fermi potentials $\varphi_\alpha$, and the mobilities $\mu_\alpha$. It is

$$\boldsymbol{j}_\alpha = z_\alpha^2 q \mu_\alpha n_\alpha \nabla \varphi_\alpha \quad \text{with} \quad \alpha \in \{\mathrm{n}, \mathrm{p}, \mathrm{x}\}. \tag{2}$$

The charge carrier densities $n_\alpha$ are also expressed as functions of the quasi-Fermi potentials (with $\alpha \in \{\mathrm{n}, \mathrm{p}, \mathrm{x}\}$)

$$n_\alpha(\psi, \varphi_\alpha) = N_\alpha \mathcal{F}_\alpha(\eta_\alpha) \text{ with } \eta_\alpha = \frac{q_\alpha(\varphi_\alpha - \psi) + z_\alpha E_{\alpha,0}}{k_{\mathrm{B}} T}, \tag{3}$$

with the effective density of state $N_\alpha$ of the respective carrier type in parabolic approximation, the temperature $T = 300$ K, and the Boltzmann constant $k_{\mathrm{B}}$. The intrinsic conduction and valence band edges $E_{\mathrm{c},0} = E_{\mathrm{n},0} = -\chi_{\mathrm{e}}$ and $E_{\mathrm{v},0} = E_{\mathrm{p},0} = -\chi_{\mathrm{e}} - E_{\mathrm{g}}$ are related to the electron affinity $\chi_{\mathrm{e}}$ and the band gap $E_{\mathrm{g}}$, and the intrinsic vacancy energy level $E_{\mathrm{x},0} = -\chi_{\mathrm{e}} - \delta_{\mathrm{x}}$ is expressed using a vacancy energy offset $\delta_{\mathrm{x}}$. For the carrier statistics functions $\mathcal{F}_\alpha$, we use the Fermi-Dirac integral of order 1/2 for electrons and holes ($\mathcal{F}_{\mathrm{n}}$ and $\mathcal{F}_{\mathrm{p}}$) and the Fermi-Dirac integral of order -1 ($\mathcal{F}_{\mathrm{x}}$) for vacancies, to consider their drift and different nonlinear diffusion from the volume exclusion effect [86]. These equations are coupled to Poisson's equation via the electrostatic potential $\psi$

$$-\nabla \cdot (\varepsilon_0 \varepsilon_{\mathrm{r}} \nabla \psi) = \sum_{\alpha \in \{\mathrm{n,p,x,d}\}} q_\alpha n_\alpha, \tag{4}$$

with the electric vacuum permittivity $\varepsilon_0$, the relative permittivity $\varepsilon_{\mathrm{r}}$, a donor type background doping density $n_{\mathrm{d}}$ and the charge $q_{\mathrm{d}} = q z_{\mathrm{d}}$ with $z_{\mathrm{d}} = 1$. At the two metal-semiconductor contacts at the locations $x = x_{\mathrm{L}}$ (left contact) and $x = x_{\mathrm{R}}$ (right contact) we impose modified Schottky boundary conditions [42]

$$\psi(\boldsymbol{x}) = \phi_0(x)/q_{\mathrm{n}} + \psi_{\mathrm{a}}(x) + \Delta\psi(x), \tag{5}$$

with the intrinsic Schottky barriers $\phi_{0,\mathrm{L}} = \phi_0(x_{\mathrm{L}})$, $\phi_{0,\mathrm{R}} = \phi_0(x_{\mathrm{R}})$ and the applied electrostatic potentials $\psi_{\mathrm{a}}(x)$ at the left and right contacts. The additional term $\Delta\psi(x)$ describes a semiclassical modification to include the effect of image-charge-induced Schottky barrier lowering obtained from self-consistently solving for an additional auxiliary electrostatic potential $\psi_{\mathrm{r}}$ representing the electrostatic potential in the device without image-charge effects [41,42]. The final equation system is solved for the three quasi-Fermi potentials $\varphi_\alpha$, $\alpha \in \{\mathrm{n,p,x}\}$, and the two electrostatic potentials $\psi$, and $\psi_{\mathrm{r}}$ as unknowns. During the simulations, we set the left contact potential to ground $\psi_{\mathrm{a}}(x_{\mathrm{L}}) = 0$ V and apply a time dependent voltage $V(t) = \psi_{\mathrm{a}}(x_{\mathrm{R}}) - \psi_{\mathrm{a}}(x_{\mathrm{L}}) = \psi_{\mathrm{a}}(x_{\mathrm{R}})$ via the potential at the right electrode, always starting from initial equilibrium.

### 4.2 Dataset Generation and Preprocessing

The 2D, 4D, 5D, 7D, and 9D datasets used in Section 3.2 were generated by varying different numbers of input parameters (parameters 1-9, Table 2) while keeping the remaining parameters constant. Table 2 summarizes the variation range of each parameter and its corresponding constant value when not varied. Specifically, parameters 1-2 were varied for the 2D dataset, 1-4 for the 4D dataset, 1-5 for the 5D dataset, 1-7 for the 7D dataset, and 1-9 for the 9D dataset. For the 5D parametric dataset used in Sections 3.3-3.5, generated without varying the voltage sequence input, parameters 5-9 were varied over the intervals $\delta_{\mathrm{x}} = [0.28, 0.32]$ eV, $\mu_{\mathrm{n}} = [10^{-4}, 10^{-2}]\,\mathrm{m^2/(Vs)}$, $\mu_{\mathrm{x}} = [10^{-15}, 10^{-13}]\,\mathrm{m^2/(Vs)}$, $\phi_{0,\mathrm{L}} = [0.05, 0.17]$ eV , and $\phi_{0,R} = [0.05, 0.17]$ eV with constant values of $t_{\mathrm{p}} = 8$ s, $V_{\mathrm{max}} = 10$ V, $L = 1$ µm, $n_{\mathrm{d}} = 10^{21}$ $\mathrm{m^{-3}}$. For all simulations, we assume a band gap of $E_{\mathrm{g}} = 1.3$ eV, and electron affinity of $\chi_{\mathrm{e}} = -4$ V, a relative electric permittivity and relative image-charge electric permittivity of $\varepsilon_{\mathrm{r}} = 10$.

To generate the parameter sets, each range was discretized into a grid of 10,000 values. Logarithmic grid spacing was applied to the mobilities $\mu_{\mathrm{x}}$, and $\mu_{\mathrm{n}}$ as well as the doping density $n_{\mathrm{d}}$, while linear spacing was used for all other parameters. For each simulation, parameter values were selected independently using a pseudo-random number generator. Each simulation produced the current response $I(t)$ associated with the corresponding voltage function $V(t)$ and model parameters. For high-throughput dataset generation, we distributed the simulations on three CPU systems to achieve approximately 360 simulations/h translating to 8640 simulations/day.

For input to the ANN models, the $I(t)$ and $V(t)$ sequences were linearly interpolated and resampled to obtain equally spaced time steps. The time step $\Delta t$ was chosen to be as large as possible without significantly affecting the validation error due to quantization effects. For the 9D dataset, $\Delta t = 10$ ms was used, while for the 5D parametric dataset, generated without varying the voltage sequence input, a larger step size of $\Delta t = 50$ ms was selected due to the fixed voltage period. After resampling, scaling was applied as described in the following subsection.

**Table 2.** Parameter ranges and constant values used for generating the datasets. Parameters 1-2 were varied for 2D, 1-4 for 4D, 1-5 for 5D, 1-7 for 7D, and 1-9 for 9D. For the 5D parametric dataset used in Sections 3.3-3.5, generated without varying the voltage input sequence, parameters 5-9 were varied, with $t_\mathrm{p} = 8$ s, $V_\mathrm{max} = 10$ V, $L = 1$ μm, $n_\mathrm{d} = 10^{20}$ m$^{-3}$ fixed.

| No. | Parameter | Range | Constant Value | Unit |
|---|---|---|---|---|
| 1 | $t_\mathrm{p}$ | [1, 10] | - | s |
| 2 | $V_\mathrm{max}$ | [1, 10] | - | V |
| 3 | $L$ | [1, 2] | 1 | μm |
| 4 | $n_\mathrm{d}$ | $[10^{19}, 10^{21}]$ | $10^{20}$ | m$^{-3}$ |
| 5 | $\delta_\mathrm{x}$ | [0.28, 0.32] | 0.3 | eV |
| 6 | $\mu_\mathrm{n}$ | $[10^{-4}, 10^{-2}]$ | $1.3 \times 10^{-3}$ | m²/(Vs) |
| 7 | $\mu_\mathrm{x}$ | $[10^{-14}, 10^{-12}]$ | $8.5 \times 10^{-14}$ | m²/(Vs) |
| 8 | $\phi_{0,\mathrm{L}}$ | [0.05, 0.17] | 0.167 | eV |
| 9 | $\phi_{0,\mathrm{R}}$ | [0.05, 0.17] | 0.148 | eV |

### 4.3 Input-Output Scaling

Before training, all input features are scaled to the interval [0,1] using min-max normalization based on the minimum and maximum values across the dataset. This applies to both the voltage sequence $V(t)$ and the time-independent input parameters $x_\mathrm{n}$ with $x \in \{L, n_\mathrm{d}, \delta_\mathrm{x}, \mu_\mathrm{n}, \mu_\mathrm{x}, \phi_{0,\mathrm{L}}, \phi_{0,\mathrm{R}}\}$. Scalar input parameters are expanded along the time dimension to match the length of the voltage sequence, forming a multichannel input sequence.

To define a consistent scaling of the output current sequences $I(t)$ across the design space, we first extract the maximum current $I_{\mathrm{max}}$ from each current sequence in the training dataset and fit a log-normal distribution to the resulting set of maxima. The fitted distribution provides a global mean $\bar{\mu}_{\mathrm{N}}$ and standard deviation $\bar{\sigma}_{\mathrm{N}}$ in logarithmic scale, which are used to standardize the individual current maxima as

$$I_{\mathrm{max,N}} = \frac{\log_{10}(I_{\mathrm{max}}) - \bar{\mu}_{\mathrm{N}}}{\bar{\sigma}_{\mathrm{N}}}. \tag{6}$$

Each current sequence $I(t)$ is then resampled and scaled using its original maximum, the standardized maximum value, and the mean $\bar{\mu}_{\mathrm{max,N}}$ of all standardized maximum values, according to

$$I_{\mathrm{n}} = I(t) \times \frac{I_{\mathrm{max,N}}}{I_{\mathrm{max}}\, \bar{\mu}_{\mathrm{max,N}}}. \tag{7}$$

This transformation maps current sequences with initially log-normally distributed maxima to a set of sequences with reduced variance and approximately normally distributed peak values around one, while preserving the overall shape and dynamics of the current response. In particular, preserving the original structure of positive and negative current regions ensures balanced error weighting during training and avoids distortions introduced by alternative offset-based scaling methods. Because the scaling factor depends only on the global distribution parameters $\bar{\mu}_{\mathrm{max,N}}$, $\bar{\mu}_{\mathrm{N}}$ and $\bar{\sigma}_{\mathrm{N}}$, predicted sequences $I_{\mathrm{n}}$ can be rescaled without access to the full dataset, by applying

$$I(t) = \frac{I_{\mathrm{n}}}{I_{\mathrm{max,n}}} \times 10^{\,\bar{\mu}_{\mathrm{max,N}} I_{\mathrm{max,n}} \bar{\sigma}_{\mathrm{N}} + \bar{\mu}_{\mathrm{N}}}, \tag{8}$$

where $I_{\mathrm{max,n}}$ is the maximum value of the predicted sequence $I_{\mathrm{n}}$.

### 4.4 ANN Architecture and Training

The ANN model is based on a sequence-to-sequence architecture comprising one or two stacked long short-term memory (LSTM) layers [61], followed by dropout regularization and a fully connected output layer. Hidden layer sizes are selected between 64 and 144 units depending on the dataset dimensionality. Models are trained using the Adam optimizer [62]with mean squared error as the loss function.

Hyperparameters, including the number of LSTM layers, hidden layer size, learning rate, batch size, and dropout rate, were selected via randomized search. The datasets were randomly partitioned into training, validation, and test subsets, using fractions of 0.8, 0.1, and 0.1, respectively, for the 2D-9D datasets, and 0.6, 0.2, and 0.2 for the 5D parametric dataset generated without varying the voltage input sequence. The validation set was used for hyperparameter tuning during training, while the test set was reserved exclusively for final performance evaluation based on the normalized mean absolute error (nMAE)

Separate ANN models are trained for each dataset. All models use one or two LSTM layers, with 64 to 144 hidden units per layer, learning rates between $1 \times 10^{-4}$ and $8 \times 10^{-4}$, batch sizes of 512, and a dropout rate of 0.1. A summary of the hyperparameter settings selected for each dataset is provided in Table 3.

**Table 3.** Overview of the hyperparameters corresponding to the results shown in Figure 4. The column labeled 5D (parametric) refers to the 5D dataset generated by varying only static parameters, with a fixed voltage sequence that was not used as an input feature.

| | 2D | 4D | 5D | 7D | 9D | 5D (parametric) |
|---|---|---|---|---|---|---|
| No. layers | 2 | 2 | 2 | 2 | 2 | 1 |
| Hidden units | 144 | 144 | 144 | 144 | 144 | 64 |
| Epochs | 46,743 | 40,528 | 29,847 | 15,053 | 19,124 | 28,477 |
| Batch size | 512 | 512 | 512 | 512 | 512 | 512 |
| Drop out ratio | 0.1 | 0.1 | 0.1 | 0.1 | 0.1 | 0.1 |
| Learn rate | $10^{-4}$ | $2 \times 10^{-4}$ | $1.5 \times 10^{-4}$ | $2 \times 10^{-4}$ | $1.5 \times 10^{-4}$ | $8 \times 10^{-4}$ |
| No. samples | $10^4$ | $10^4$ | $2 \times 10^4$ | $2 \times 10^4$ | $2 \times 10^4$ | $2 \times 10^4$ |

### 4.5 Runtime and Performance Analysis

All runtimes were measured on a workstation equipped with an Intel® Core™ i9-14900KF CPU and an NVIDIA® GeForce™ RTX 4090 GPU, using MATLAB R2024b and the Deep Learning Toolbox for inference on either the CPU or GPU. For GPU-based inference, a batch size of 10,000 was used to enable parallel evaluation of input sequences. Runtime measurements for a single I-V curve were based on randomly sampled parameter combinations, with 1,000 finite volume simulations and either 10,000 (CPU) or 1,000,000 (GPU) ANN forward evaluations of the model trained on the 5D parametric dataset. Reported runtimes represent averages over these measurements. For ANN inference, the first ten forward evaluations (CPU) or the first ten batches (GPU) were excluded from averaging to account for initialization overhead, which adds a constant delay (< 1 s) independent of task size. The other ANN runtimes reported in Table 1 for CPU inference correspond to the measured execution times for each task using the ANN surrogate model. In contrast, the runtimes for the finite volume model and the GPU-based metric mapping and sensitivity analysis tasks are estimated by multiplying the measured runtime for a single I-V simulation by the number of model evaluations used for each task.

Performance is assessed using a normalized mean absolute error (nMAE) between predicted and ground truth current sequences. For each test sample, the scaled ground truth current $I_{\mathrm{n,truth}}(t)$ and the corresponding scaled predicted current $I_{\mathrm{n,pred}}(t)$ are first divided by the global maximum absolute value $\max(|I_{\mathrm{n,truth}}|, |I_{\mathrm{n,pred}}|)$ of $I_{\mathrm{n,truth}}(t)$ and $I_{\mathrm{n,pred}}(t)$. The nMAE is then computed as half the mean absolute difference between the normalized sequences

$$\mathrm{nMAE} = \frac{1}{2N}\sum_{i=1}^{N}\left|\frac{I_{\mathrm{n,pred}}(t_i) - I_{\mathrm{n,truth}}(t_i)}{\max(|I_{\mathrm{n,truth}}|, |I_{\mathrm{n,pred}}|)}\right|, \tag{9}$$

where $N$ is the number of time points in the sequence, and $t_i$ is the ith element of the time vector. This definition strictly constrains nMAE values to the interval [0, 1].

### 4.6 Metric Definitions

To quantify specific features of the I-V response, we define two normalized scalar metrics derived from the simulated current–voltage sequences. For the constrained and normalized on-off ratio $R$, we define

$$R = \frac{|I_{\mathrm{on}}|}{|I_{\mathrm{off}}| + |I_{\mathrm{on}}|}, \tag{10}$$

where $I_{\mathrm{on}}$ is the current at $V = 5$ V in the upsweep from 0 V to $V_{\mathrm{max}}$ and $I_{\mathrm{off}}$ is the current at $V = 5$ V in the downsweep from $V_{\mathrm{max}}$ back to 0 V. This definition ensures that $0 < R < 1$ resulting in clockwise hysteresis in the right branch for $0.5 < R < 1$ and counterclockwise hysteresis for $0 < R < 0.5$.

To quantify the symmetry of a simulated I-V curve, we first apply min-max scaling to the current values such that the scaled current lies between 0 and 1. The symmetry metric $S$ is then defined as

$$S = 1 - \Delta I_{\mathrm{max}}, \tag{11}$$

where $\Delta I_{\mathrm{max}} = |I(V_{\mathrm{max}}) - I(V_{\mathrm{min}})|$ is the absolute difference between the scaled current values at the maximum applied voltage $V_{\mathrm{max}} > 0$ and the minimum voltage $V_{\mathrm{min}} < 0$ during the sweep. The resulting symmetry value $S$ is constrained to the interval [0,1], with $S = 1$ indicating perfect symmetry.

### 4.7 Sensitivity Analysis

Variance-based sensitivity analysis was carried out using the Sobol' decomposition [72,73] with a total of 170,000 parameter combinations. The analysis was performed using UQLab [87,88], with Sobol sampling across the entire parameter range of the 5D parametric dataset. Device metrics were obtained by predicting I-V curves with the ANN model trained on the 5D parametric dataset.

### 4.8 ANN-Based Optimization Tasks

For the inverse modeling task in Section 3.3, the cost function $\mathcal{L}$ was defined as the mean logarithmic percentage error between the magnitudes of the ANN-predicted current sequence $I_{\mathrm{pred}}$ and the measured current sequence $I_{\mathrm{meas}}$, interpolated to match the sequence length. The cost function is given by

$$\mathcal{L} = \frac{1}{N}\sum_{i=1}^{N}\left|\frac{\log_{10}(|I_{\mathrm{meas}}(t_i)|) - \log_{10}\left(\left|I_{\mathrm{pred}}(t_i)\right|\right)}{\log_{10}(|I_{\mathrm{meas}}(t_i)|)}\right|, \tag{12}$$

where $N$ is the number of time points in the sequence, and $t_i$ is the ith element of the time vector.

For the single-objective optimization task in Section 3.5, the cost function was defined as $\mathcal{L} = (S - S_0)^2 + (R - R_0)^2$, where $S$ is the symmetry metric, $R$ the on-off ratio predicted by the ANN model and the target values were set to $S_0 = 0$ and $R_0 = 1$. Both metrics were constrained to the interval [0,1]. Optimization was performed using MATLAB's MultiStart framework, with 500 randomized initializations for both optimization tasks. As local nonlinear solver, fmincon was used with the Sequential Quadratic Programming (SQP) algorithm using a maximum of 500 iterations.

For the multi-objective optimization task in Section 3.5, $S$ and $R$ were treated as independent objectives, defined as $\mathcal{L}_1 = S$ and $\mathcal{L}_2 = -R$ to express the problem as a two-objective minimization. Both metrics were constrained to the interval [0,1]. Optimization was performed using MATLAB's gamultiobj solver with a population size of 200 and a maximum of 200 generations, resulting in 40,000 surrogate model evaluations. For all three tasks, the ANN model trained on the 5D parametric dataset was used. The five model parameters were used in their scaled form $x_{\mathrm{n}}$ with $x \in \{\delta_{\mathrm{x}}, \mu_{\mathrm{n}}, \mu_{\mathrm{x}}, \phi_{0,\mathrm{L}}, \phi_{0,\mathrm{R}}\}$ and were constrained to the bounds used during ANN training. All optimization tasks were implemented in MATLAB R2024b (The MathWorks, Inc., Natick, MA, USA). The model parameters obtained from the inverse modeling task described in Section 3.3, corresponding to the I-V curves shown in Figure 5a,b, are summarized in Table 4. All optimization tasks were performed using the second voltage cycle to ensure representative device behavior.

**Table 4** Model parameters obtained from inverse modeling of the experimental I-V curves shown in Figure 5a,b.

| **Parameter** | **Set 1 (Figure 5a)** | **Set 2 (Figure 5b)** | **Unit** |
|---|---|---|---|
| $t_{\mathrm{p}}$ | 8 | 8 | s |
| $V_{\mathrm{max}}$ | 10 | 10 | V |
| $L$ | 1 | 1 | µm |
| $n_{\mathrm{d}}$ | $10^{21}$ | $10^{21}$ | $\mathrm{m}^{-3}$ |
| $\delta_{\mathrm{x}}$ | 0.28 | 0.28 | eV |
| $\mu_{\mathrm{n}}$ | 0.00196 | 0.0028 | $\mathrm{m}^2/(\mathrm{Vs})$ |
| $\mu_{\mathrm{x}}$ | $5.6829 \times 10^{-14}$ | $1.0462 \times 10^{-13}$ | $\mathrm{m}^2/(\mathrm{Vs})$ |
| $\phi_{0,\mathrm{L}}$ | 0.0500 | 0.1236 | eV |
| $\phi_{0,\mathrm{R}}$ | 0.0503 | 0.1085 | eV |

## Data Availability

The codes and data supporting the current study are available from the corresponding authors upon reasonable request.

## Acknowledgments

The research was funded by the German Research Foundation through the Collaborative Research Centre CRC 1461, the Carl Zeiss Foundation via the project Memwerk. This work was partially supported by the Leibniz competition 2020 (NUMSEMIC, J89/2019), and the Deutsche Forschungsgemeinschaft (DFG, German Research Foundation) under Germany's Excellence Strategy – The Berlin Mathematics Research Center MATH+ (EXC-2046/1, project ID: 390685689)

## Author Contributions

B.S. conceptualized the project and developed the neural network model. B.S. and E.S. performed the simulations and visualized the results. B.S., D.A., and P.F. developed the charge transport model. All authors discussed the results and contributed to the writing of the manuscript.

## Competing Interests

The authors declare no competing interests.